# Performance Evaluation of Mobile U-Navigation based on GPS/WLAN Hybridization

*1,Corresponding Author Wan Mohd Yaakob Wan Bejuri, [2]Mohd Murtadha Mohamad, [3]Maimunah Sapri, [4]Mohd Adly Rosly
*[1,2,4] Faculty of Computer Science & Information Systems, Universiti Teknologi Malaysia*
*[2]Centre of Real Estate Studies, Universiti Teknologi Malaysia*
*[1]wanmohdyaakob@ieee.org, [2]murtadha@utm.my, [3]maimunahsapri@utm.my,*
*[4]madlysalaf@gmail.com*

***Abstract***

*This paper present our mobile u-navigation system. This approach utilizes hybridization of wireless local area network and Global Positioning System internal sensor which to receive signal strength from access point and the same time retrieve Global Navigation System Satellite signal. This positioning information will be switched based on type of environment in order to ensure the ubiquity of positioning system. Finally we present our results to illustrate the performance of the localization system for an indoor/ outdoor environment set-up.*

## 1. Introduction

The movement into the ubiquitous computing realm will integrate the advances from both mobile and pervasive computing. The idea is to surround ourselves with computers and software that are carefully tuned to offer us unobtrusive assistance as we navigate through our work and personal lives. Though these terms are often used interchangeably, they are conceptually different and employ different ideas of organizing and managing computing services [1] [2] [3]. The development of mobile computing has grown rapidly, since it is a most cool gadget that compromise mobility, efficiency and effective to the end users. Nowadays, mobile computer become multipurpose device, is not just as communicator device, but also can be navigator device. Most of the high-end mobile device has been equipped with Global Positioning System (GPS) navigation system that gives navigation services to end user if they want to travel to interest destination. However, the navigation by using GPS was suffered in obstruction environment especially user inside building (indoor environment) [4] [5] [6] [7] [8] [9] [10] [11] [12] [13].In addition, the object such as tree, high building, high wall and also people walking might be the contributors of the obstruction. These obstructions sometimes moved to another location which usually occurred in indoor environment and finally make it difficult to estimate user's position. Therefore, there is a need an alternative solution in order to ensure it also can get positioning information inside building (For example situation: a visitor want to find her friend in a complex building office) The hybridization of positioning method may tend to solve this issue in order to determine positioning information with more pervasive, reliable and ubiquity.

Conventional method using hybridization between GPS and mechanical sensor (for example: inertial navigation system) are known quite successful in term of navigation either inside and outside building, this solution are not tend very much to solve this issue since it may make end users feel harsh on device integration and configuration. The mobility issue also becomes to be major problem too. To ensure the device can be more mobility, there is a need to hybrid GPS with internal sensor (WLAN or, Camera or Bluetooth) within a mobile device itself. In this paper, we are focusing the mobile u-navigation system development in term to how to get positioning information in both of inside building and also outdoor environment, by using hybridization of GPS and WLAN. The structure of the paper is as follows. Section 2 will present the reviews related work to mobile U-navigation system. Section 3 will present an overview of our proposed method. The details of our result are covered in section 4. Finally, conclusions are given in section 5.

## 2. Related Work

The main issue of most popular mobile navigation system; (GPS standalone), is its positioning accuracy. Furthermore. that technology also making the navigation system as overall become difficult to be operated in obstructed environment; especially inside building. Therefore, the concept of indoor positioning system has been introduced. The world first indoor WLAN positioning; known as RADAR [14], which adopts the nearest neighbour(s) in signal-space technique. The accuracy of the proposed system is around 2–3 m. Later [15], RADAR was enhanced by a Viterbi-like algorithm. Its result is that the 50 percent of the RADAR system is around 2.37–2.65 m and its 90 percentile is around 5.93–5.97 m. Horus system [16], [17] offered a joint clustering technique for location estimation, which uses the probabilistic method described previously. Each candidate location coordinate is regarded as a class or category. The experiment results show that this technique can acquire an accuracy of more than 90% to within 2.1 m. A grid-based Bayesian location-sensing system over a small region of their office building [18], achieving localization and tracking to within 1.5 m over 50% of the time. Nibble [19], one of the first systems of this generation, used a probabilistic approach (based on Bayesian network) to estimate a device's location. In [20], Battiti et al. proposed a location determination method by using neural-network-based classifier by adopted multilayer perceptron (MLP) architecture and one-step secant (OSS) training method. They reported that only five samples of signal strengths in different locations are sufficient to get an average distance error of 3 m. Increasing the number of training examples helps decrease the average distance error to 1.5 m.

At the same time, the research goes beyond by making positioning system more pervasive and ubiquity, the company known as Qualcomm overcome the limitations of conventional GPS, and provide GPS indoors technique with an average of 5–50 m accuracy in most indoor environments. A-GPS technology uses a location server with a reference GPS receiver that can simultaneously detect the same satellites as the wireless handset (or mobile station) with a partial GPS receiver, to help the partial GPS receiver find weak GPS signals. The wireless handset collects measurements from both the GPS constellation and the wireless mobile network. These measurements are combined by the location server to produce a position estimation. The Locata Corporation finally invented a new precision indoor/outdoor positioning technology called Locata [21]. It is consists of a time-synchronized pseudolite transceiver called a LocataLite which transmits GPS-like signals that allow single-point positioning using carrier-phase measurements. The result shows that it is possible to archive distance error within sub-centimeter [21]. Although the u-navigation is not new, but the technique above is more focus to the new device development, rather than utilize mobility function on mobile device.

## 3. Methodology

Generally our system is to determine positioning information inside and outside building, by using mobile device. In order to utilize mobility function in our system, the hybridization of internal positioning sensor (such as: GPS, GSM, WLAN and Bluetooth) within a mobile device is needs. Refer to Figure 1, our system generally consist of three (3) subsystems which are named as field subsystem, interface subsystem, and database subsystem. In the normal situation GPS satellite (in earth orbit) and WLAN access point (inside building) will continuously broadcasting their signal within coverage. Any mobile device that equipped WLAN and GPS sensor within their coverage will received signal. The signal will be processed by conventional indoor/outdoor positioning (as shown on Figure 2). This algorithm is aimed to switch suitable positioning method in different environment. The method indoor positioning will be started as soon as outdoor positioning sensor cannot get signal input. In our case, we are using GPS as outdoor positioning sensor, meanwhile WLAN as indoor positioning sensor. in order to select which positioning method will be used. Then, the positioning data in obtained signal will be extracted to compare with surveying data in database server. Finally the output of system will display the mapping location on mobile device screen.

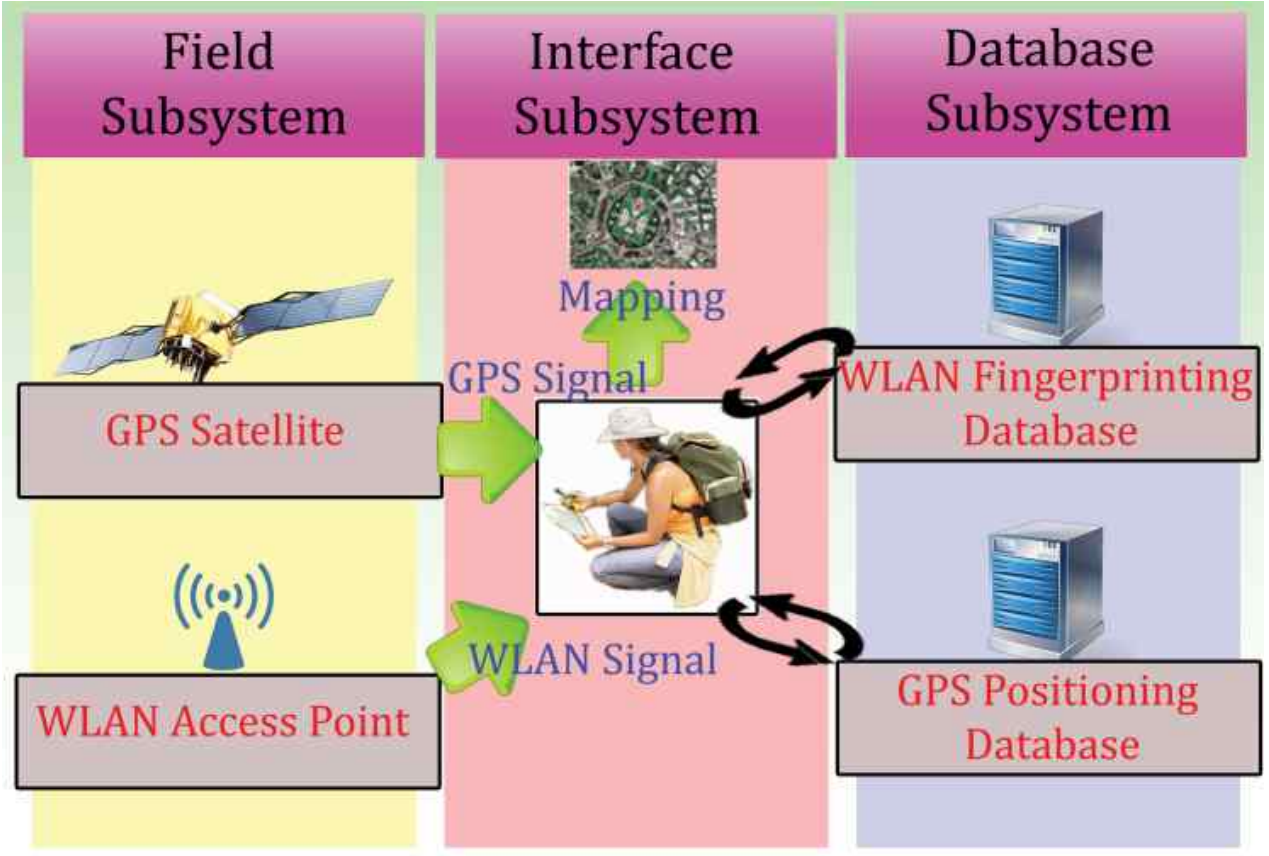

**Figure 1.** General Architecture of U-Navigation.

```
U-Navigation Algorithm{
1) Read GPS Signal;
2) If GPS Signal is valid {
A. Outdoor_Positioning ;
}
3) else {
A. Apply indoor positioning ;
}
}
```

**Figure 2.** Basic Algorithm of U-Navigation (written in Pseudocode).

### 3.1. Outdoor Positioning

In the GPS positioning we need to set up some known points or so called references points in order with compare with obtained coordinate from GPS signal. The reason behind this, is to map the real environment within mobile device screen. Thus, we prefer to use Relative-Interpolation method [22] since the coordinates of the reference point are not the absolute longitude and latitude, but the x-y coordinates relative to the (0, 0) point of the window in which the map is rendered. At each of the reference points, A and B, we need to know the coordinate (real environment) by using high precision GPS positioning on the site (for example using GPS device in Figure 4) and, longitude and latitude as well as its X and Y coordinates on the window. Let $x_{lt}$ and $x_{bng}$ be the longitude and latitude of the reference point $x$ (A or B), and $x_x$ and $x_y$ be it is $x$ and $y$ coordinates, respectively. Let C be the unknown user's current position and $c_{lt}$ and $c_{bng}$ be the GPS data measured at the user's current position. Then, we estimate the user's $x$ and $y$ coordinates on the map, $c_x$ and $c_y$respectively, with the equations below:

$$C_x = \left(\frac{C_{bn} - A_{bn}}{B_{bn} - A_{bn}}\right)(B_x - A_x) + A_x \quad (1)$$

$$C_y = \left(\frac{C_{lt} - A_{lt}}{B_{lt} - A_{lt}}\right)(B_y - A_y) + A_y \quad (2)$$

## 3.2. Indoor Positioning

In our indoor positioning method, we prefer to use a well known WLAN fingerprinting method which known as RADAR [14]. In the RADAR WLAN positioning system, there is a searching algorithm, which it is be in main part of the system, known as KNN nearest neighbor [23]. This algorithm contributes by look-up table during the off-line phase. However, WLAN signal strength also suffer in the obstruction environment since the signal will propagate and loss if there is a blockage between AP and mobile device receiver. Theoretically, the WLAN signal path loss obeys the distance power law as described below;

$$P_r(d) = P_r(d_0) - 10nlog \left(\frac{d}{d_0}\right) + X_\sigma \tag{3}$$

Where Pr is the received power; $P_r(d_0)$ is the received power at the $d_0$ (called as reference distance), $n$ is refer to path loss exponent, which indicates the rate of the path loss increases with distance. It depends on the surrounding, building type and other obstructions. And $d_0$ is the close-in reference distance (1m) and d is the distance of separation between the RF signal transmitter and receiver (The transmitter could be AP and receiver could be mobile device receiver). The term $X_\sigma$ is a zero mean Gaussian random variable with standard deviation $\sigma$. Equation (3) is modified to include Wall Attenuation Factor ($W\ AF$). The modified distance power law is given as (4),

$$P_r(d) = P_r(d_0) - 10nlog \left(\frac{d}{d_0}\right) - T * W\ AF \tag{4}$$

Where, T is number of walls between transmitter and receiver.

$$\mathrm{d} = \mathrm{e}^{\left(\frac{\mathrm{Pr}(\mathrm{d_0}) - \mathrm{Pr}(\mathrm{d}) - \mathrm{T} * \mathrm{W\ AF}}{10}\right)} \tag{5}$$

Equation (5) has been derived from equation (4). This equation is to measure the distance between the Access Point and Mobile Node. When the mobile device or node location been calculated, the distance of every devices or nodes will be calculated by using Euclidean Distance equation (6).

$$\mathrm{Distance} = \sqrt{((\mathrm{X_1} - \mathrm{X_2})^2 + (\mathrm{Y_1} - \mathrm{Y_2})^2)} \tag{6}$$

The Location Server will calculate the distance for every device in the network and compare all the distance to find out which is the nearest device from the mobile node chosen. The nearest computation method is done by nearest neighbor(s) in signal space (NNSS). The idea is to compute the distance (in signal space) between the observed set of SS measurements, $(\mathrm{ss_1}, \mathrm{ss_2}, \mathrm{ss_3})$ and the recorded SS, $(\mathrm{ss_1'}, \mathrm{ss_2'}, \mathrm{ss_3'})$ at a fixed set of locations, and then pick the location that minimizes the distance. In order to calculate based on three (3) measurements, the equation (6) can be inherit to become as equation (7) which is D is distance between observed signal and recorded signal;

$$\mathrm{D} = \sqrt{((\mathrm{ss_1} - \mathrm{ss_1'})^2 + (\mathrm{ss_2} - \mathrm{ss_2'})^2 + (\mathrm{ss_3} - \mathrm{ss_3'})^2)} \tag{7}$$

# 4. Result

The experiments was located at Universiti Teknologi Malaysia, Johor, Malaysia and generally can categorized in two (2) types which are indoor positioning and outdoor positioning. The experiment of indoor positioning was done at level 2, FSKSM building (see Figure 5) by using 3 (three) APs (access points) in two (2) locations. Meanwhile, the outdoor positioning was done at Lingkaran Ilmu road (for detail see Figure 3). As overall, indoor positioning can be categorized in two (2) phases of experiment procedure which are offline phase and online phase. The offline data collection was obtained by taking GPS coordinate point (using GPS Trimble as shown in Figure 4) for outdoor positioning and WLAN

Signal Strength (using mobile device).in four (4) orientation (for detail about orientation, please see Figure 6) for indoor positioning. Meanwhile, during online phase, data collection was done in same process except data collection for outdoor positioning must be done by using internal GPS on mobile device. Below, we will discuss more detail about performance of proposed approach.

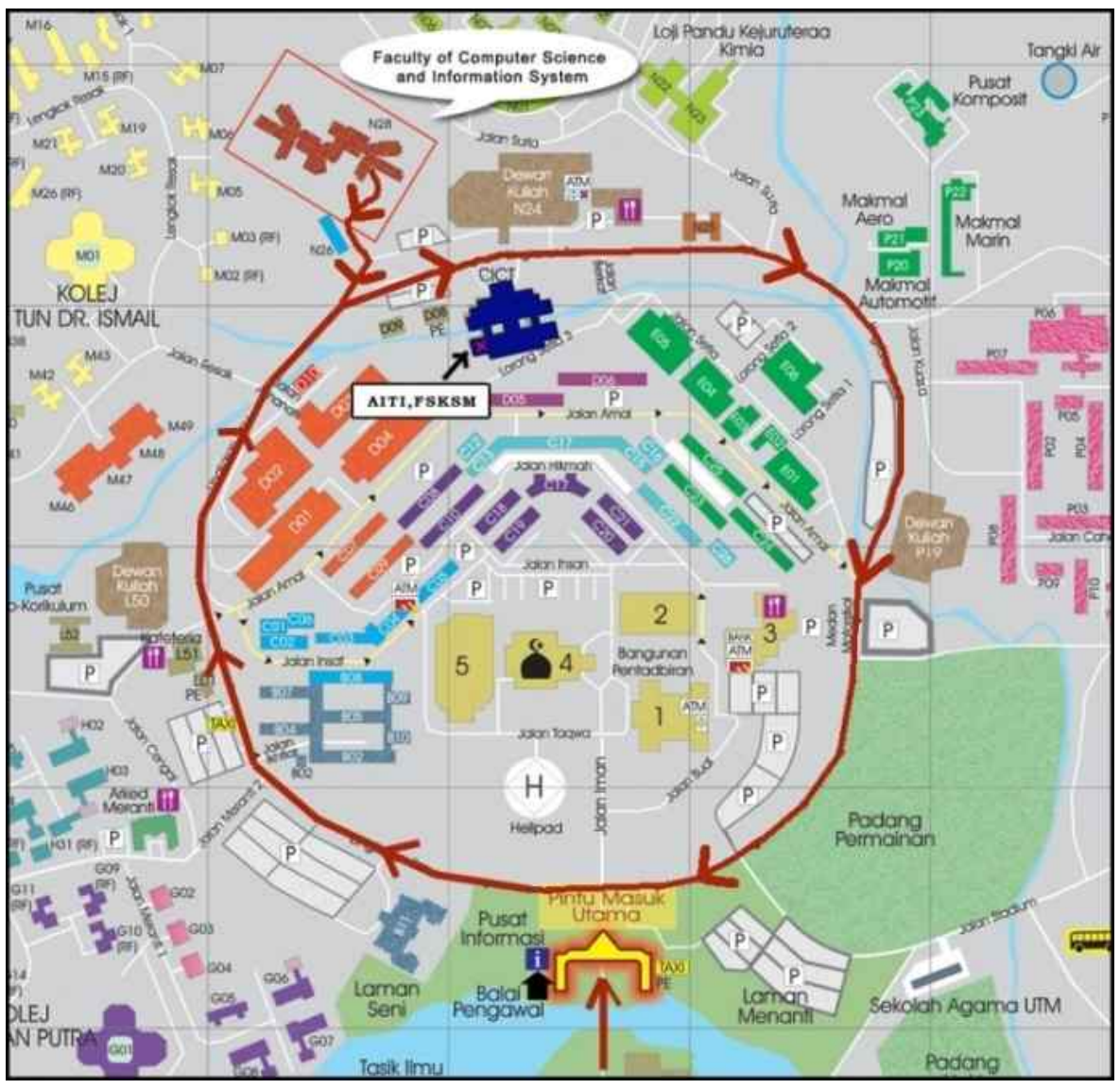

**Figure 3.** Area Covered (see red path) for Outdoor Positioning

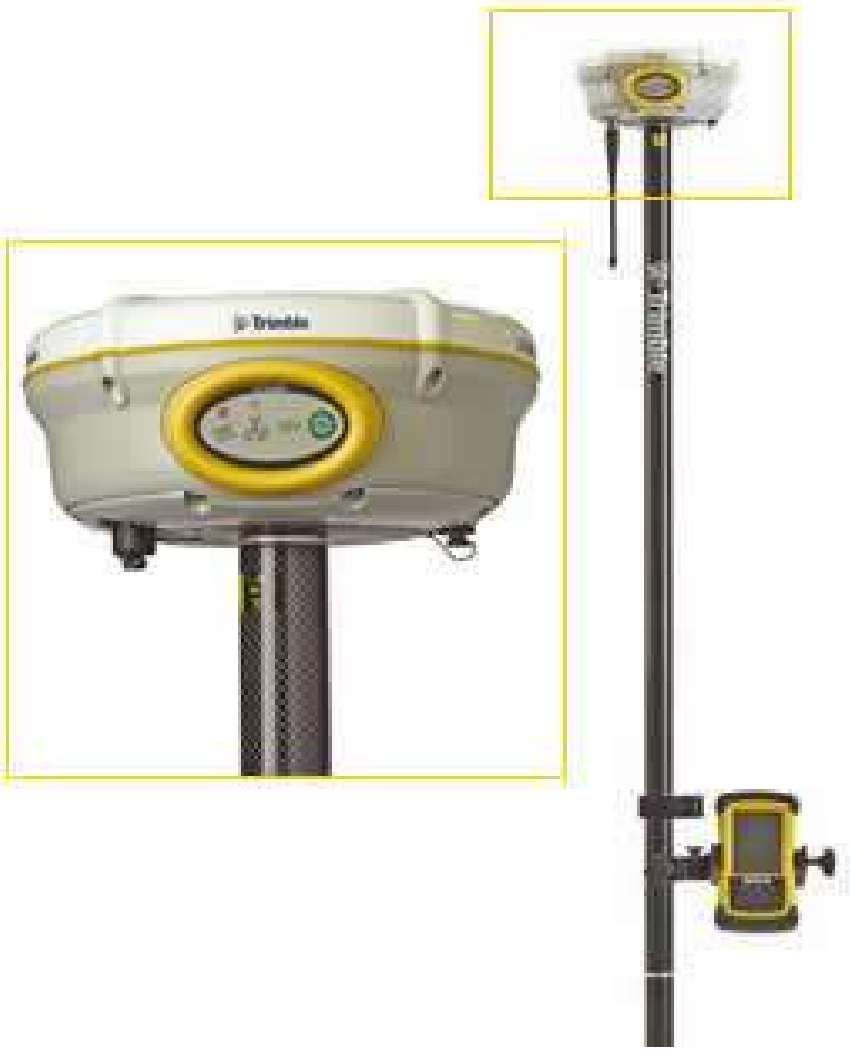

**Figure 4.** GPS Trimble Which Used for Data Collection (Outdoor Positioning)

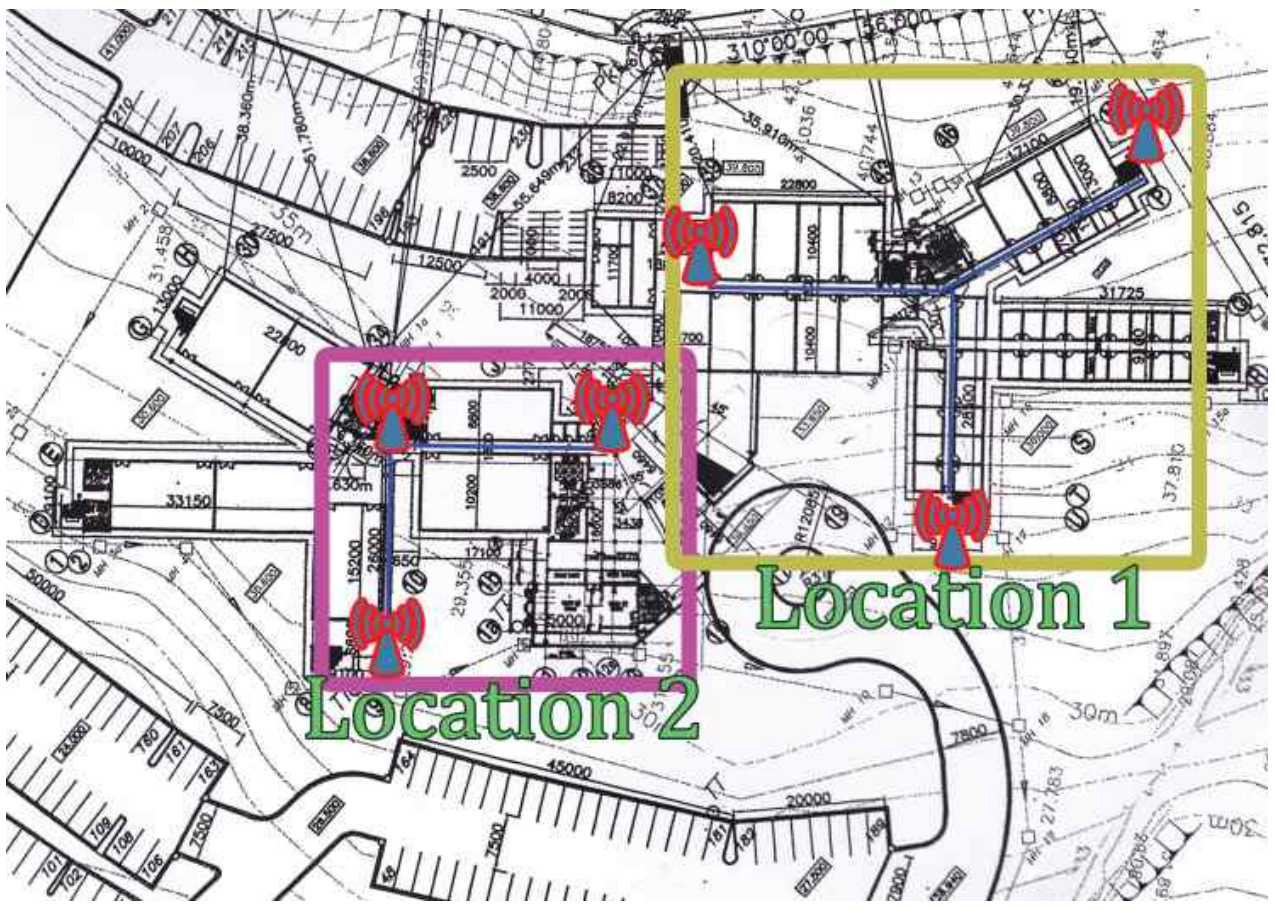

**Figure 5.** Area Covered (there are two location) for Indoor Positioning

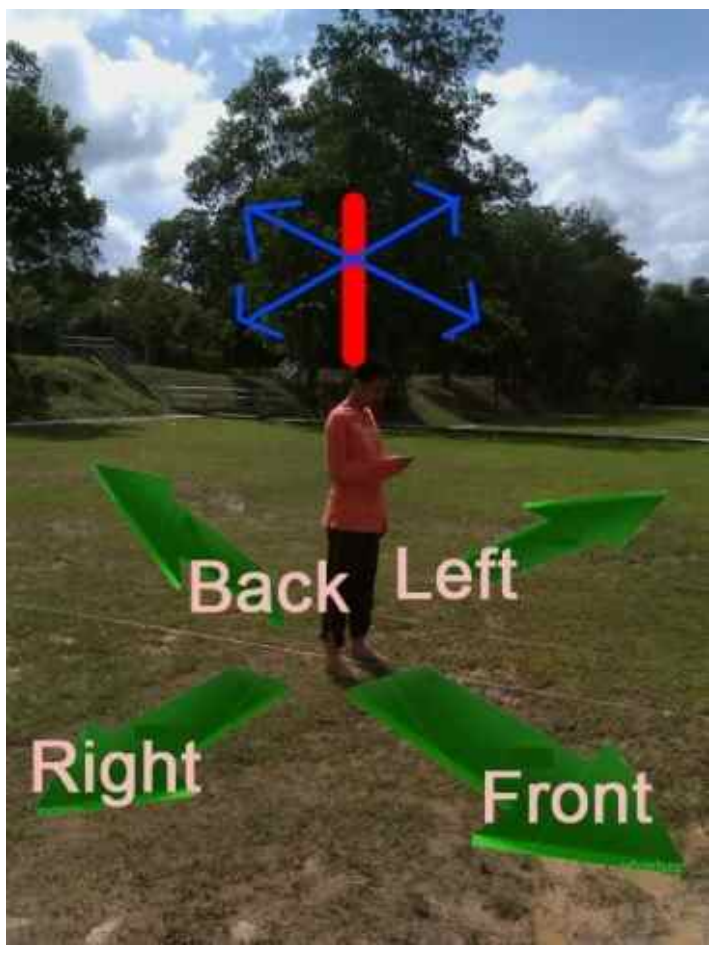

**Figure 6.** User Orientation

## 4.1. Outdoor Positioning

The measured x-y coordinates and the latitude and longitude of the reference points are shown in Table 1. We performed experiments in which the x-y coordinates of the current position obtained by clicking the mouse on the window were compared with those obtained from the outdoor positioning program 200 times and the results are summarized in Table 2. The results indicate that among the 200 experiments, on 11 occasions the error was less than 1 m, on 17 occasions the error was between 1 and 2 m, and so on. The average error was 4.875m.

**Table 1:** x,y coordinates, latitude and longitude of reference points

| | **Coordinates** | | **GPS data** | |
|---|---|---|---|---|
| | **X** | **Y** | **Latitude** | **Longitude** |
| **A** | 1842 | 1140 | 1°33'27.16"N | 103°38'11.69"E |
| **B** | 2112 | 1156 | 1°33'45.84"N | 103°38'13.82"E |

**Table 2:** Summary of the results of the outdoor positioning experiments

| **Error (m)** | 0~1 | **1~2** | **2~4 `** | **4~6** | **6~8** | **8~** |
|---|---|---|---|---|---|---|
| **Occurrence** | 11 | 17 | 61 | 51 | 33 | 27 |
| **Probability** | 5.5% | 8.5% | 30.5% | 25.5% | 16.5% | 13.5% |
| **Average Error = 4.875 m** | | | | | | |

## 4.2. Indoor Positioning

In the indoor positioning, we can see Figure 7, 8, 9 and 10 shows the RSSI strength compare to the distance between the mobile node and the access points in different of user orientation. As we can see from that figure, the trend of RSSI shows normal, which that the value increases if the distance decreases and that trend almost same in Figure 11, 12, 13 and 14. The wide difference in signal strengths between points at similar distances is explained as follows: the layout of the rooms in the building, the placement of base stations, and the location of the mobile user all have an effect on the received signal.

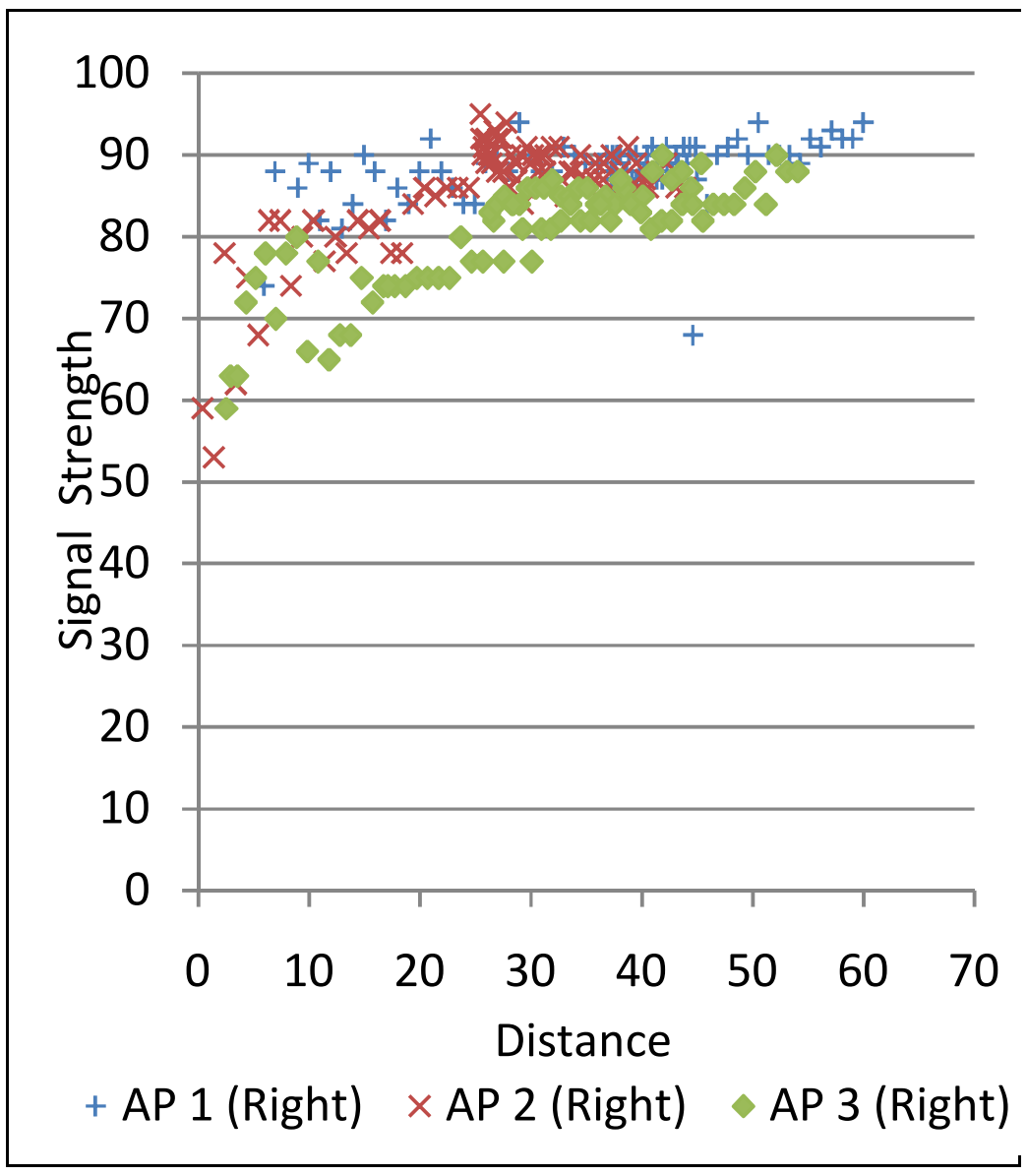

**Figure 7.** Distance and RSSI during Data Collection in Location 1 (User Orientation: 0°).

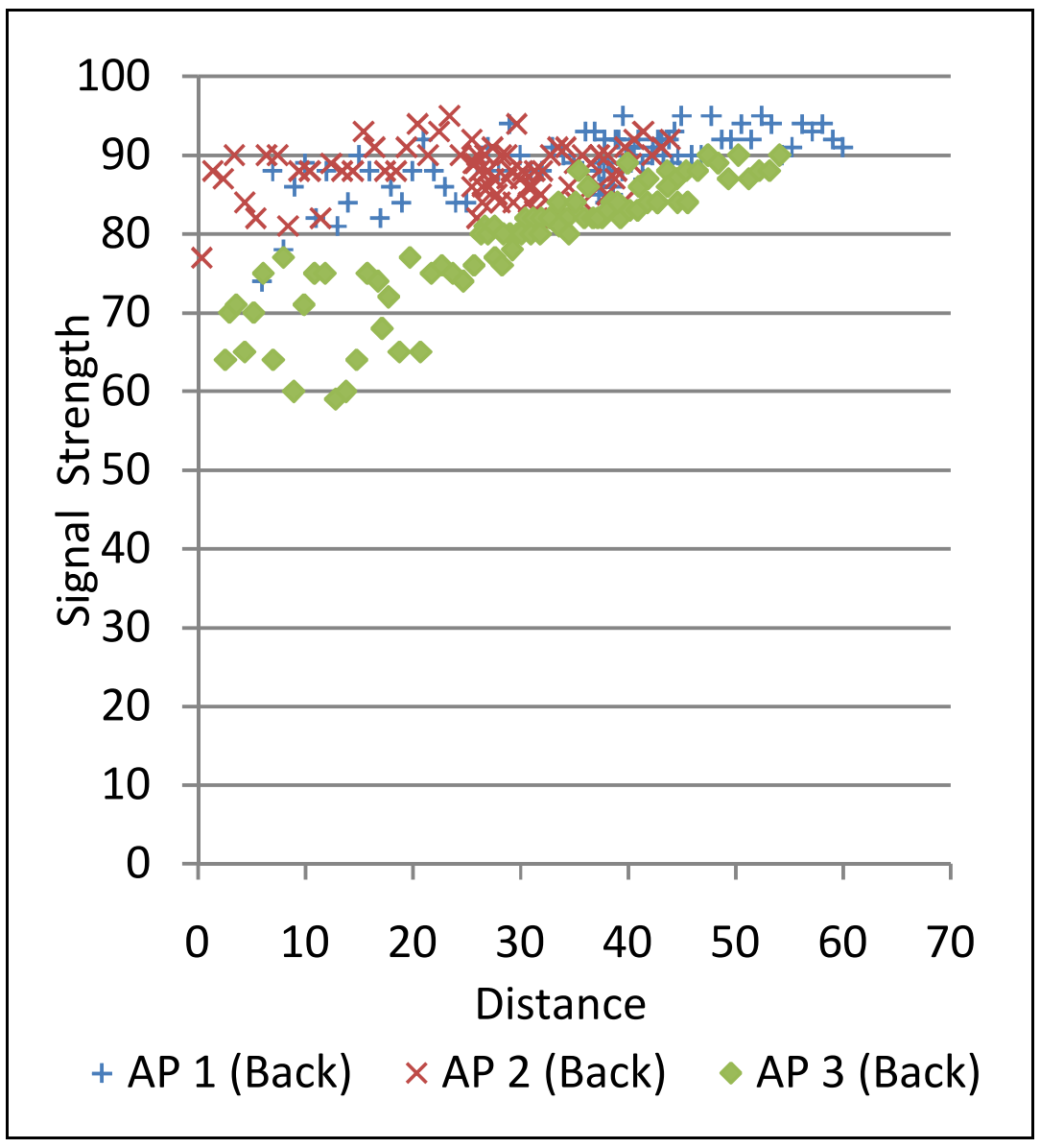

**Figure 8.** Distance and RSSI during Data Collection in Location 1 (User Orientation: 90°). (Note: Signal Strength in –dBm and Distance in meter).

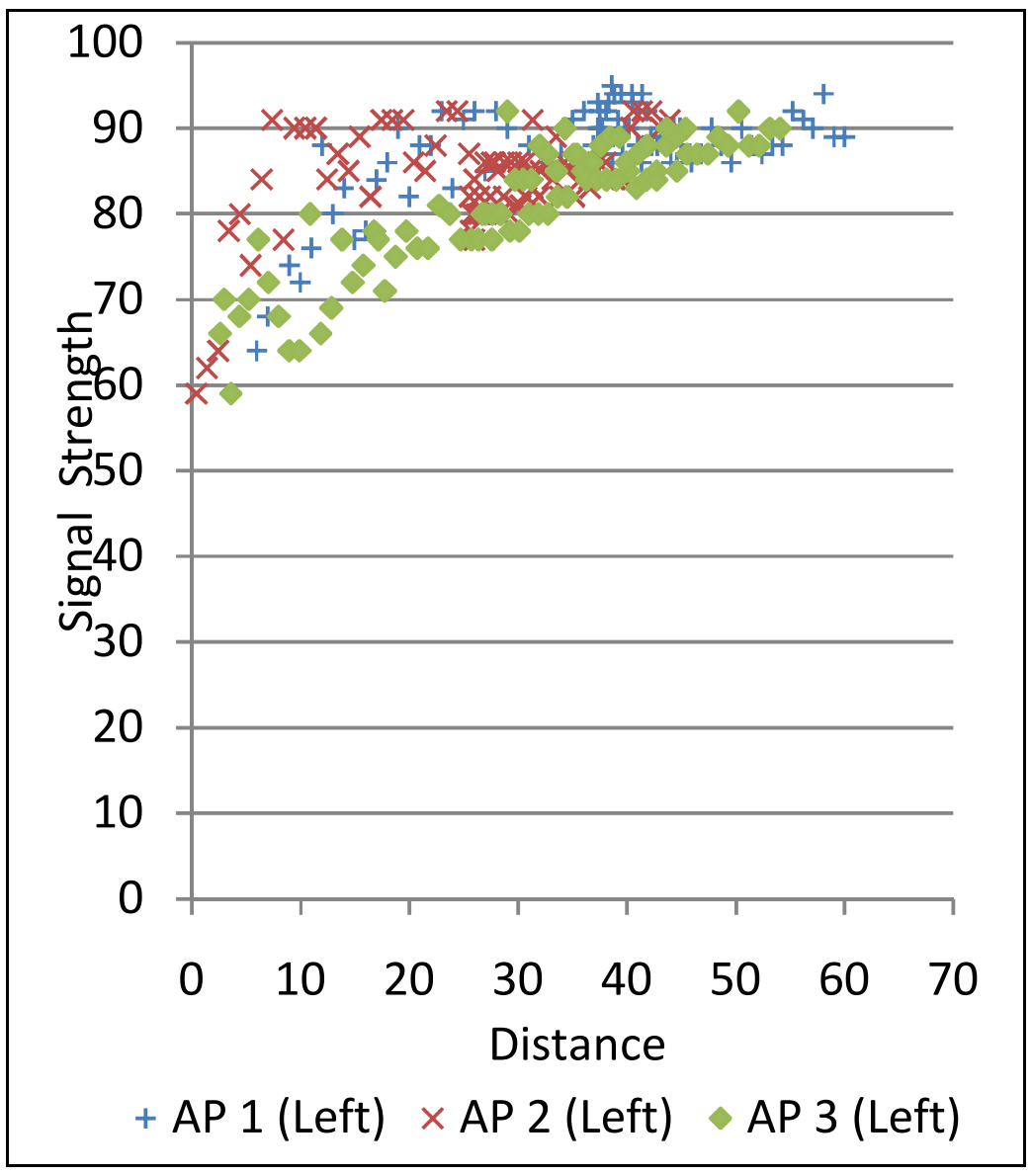

**Figure 9.** Distance and RSSI during Data Collection in Location 1 (User Orientation: 180°). (Note: Signal Strength in –dBm and Distance in meter).

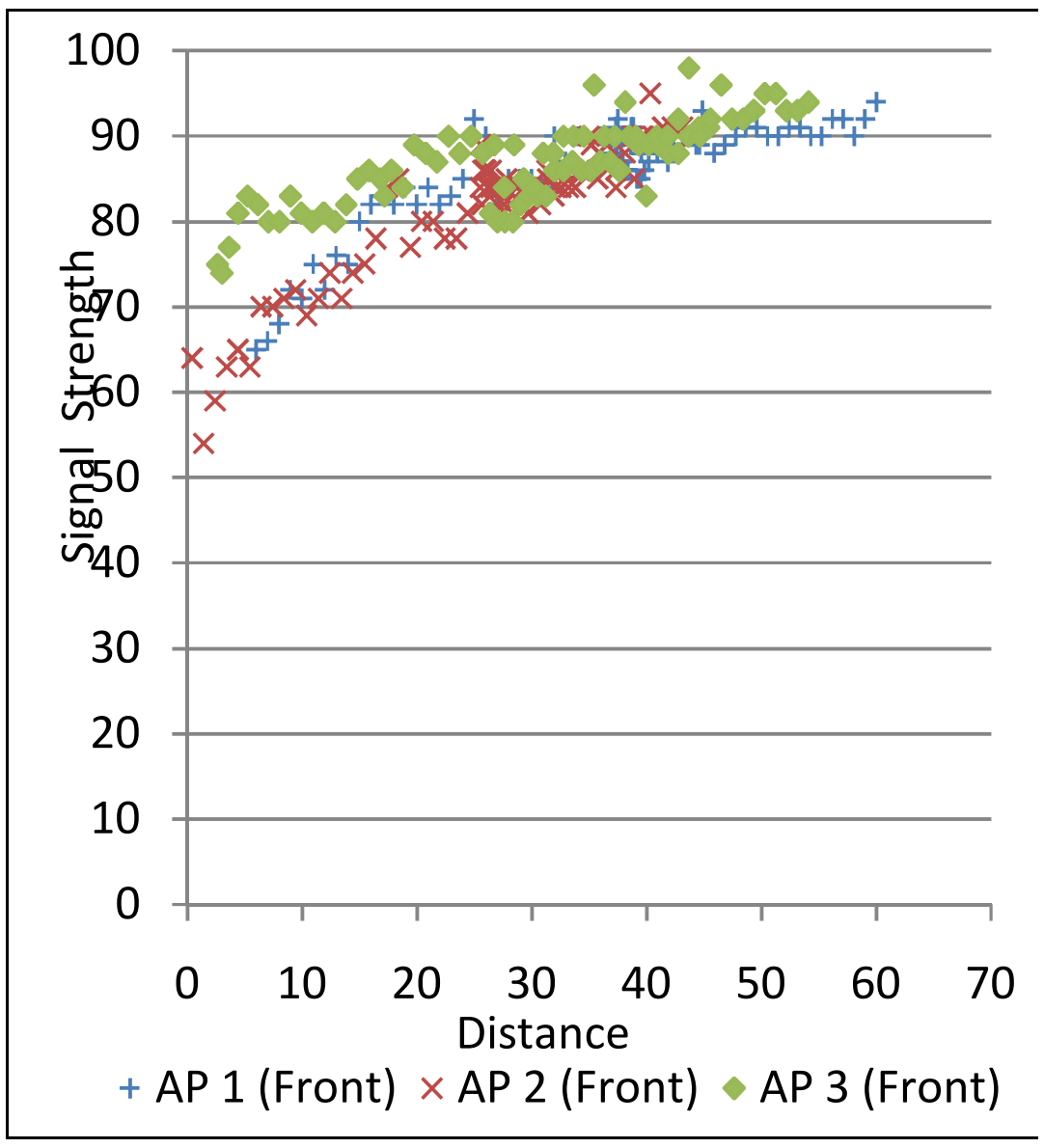

**Figure 10.** Distance and RSSI during Data Collection in Location 1 (User Orientation: 270°). (Note: Signal Strength in –dBm and Distance in meter).

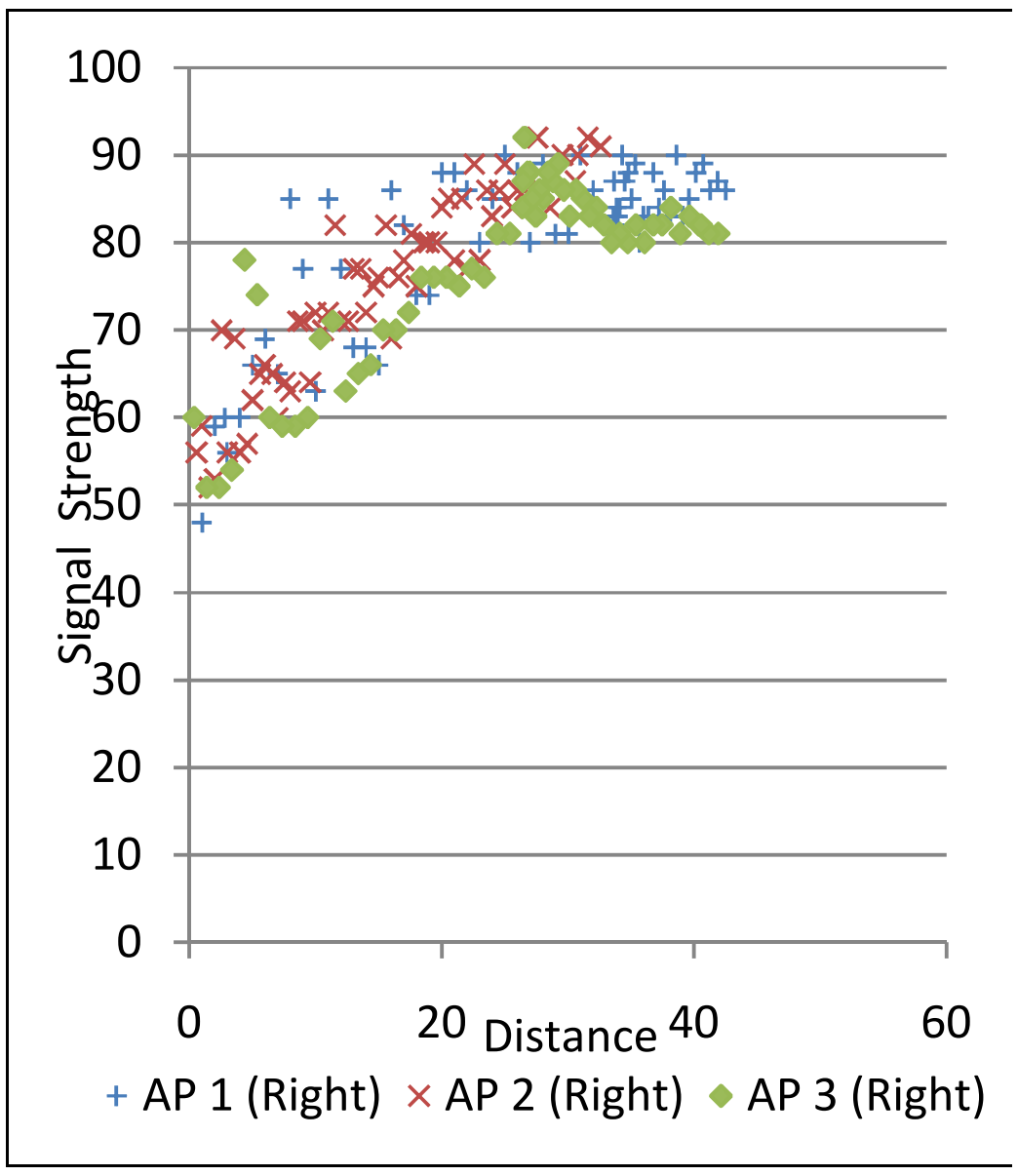

**Figure 11.** Distance and RSSI during Data Collection in Location 2 (User Orientation: $0^o$) (Note: Signal Strength in –;dBm and Distance in meter).

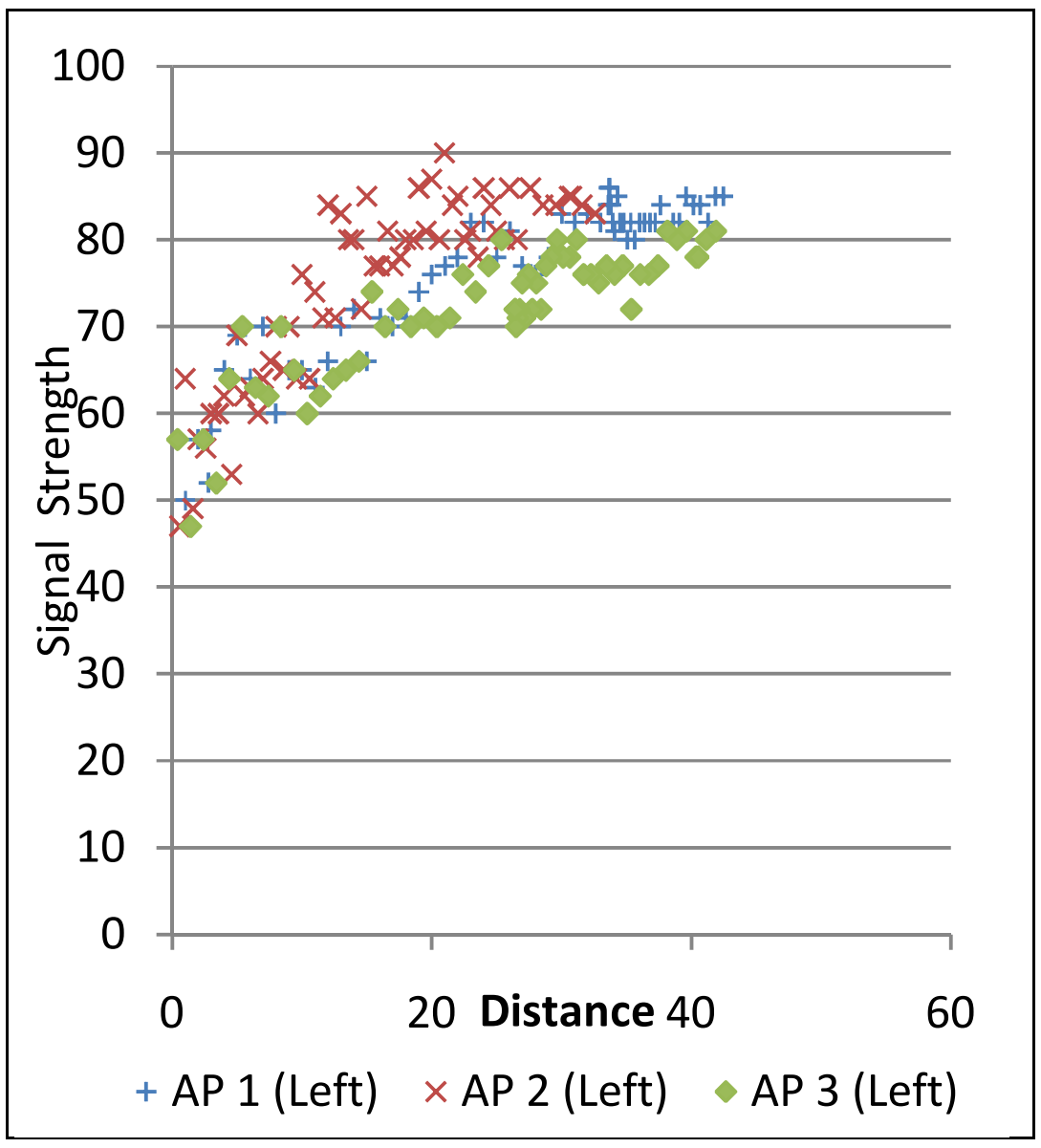

**Figure 12.** Distance and RSSI during Data Collection in Location 2 (User Orientation: $180^o$). (Note: Signal Strength in –dBm and Distance in meter).

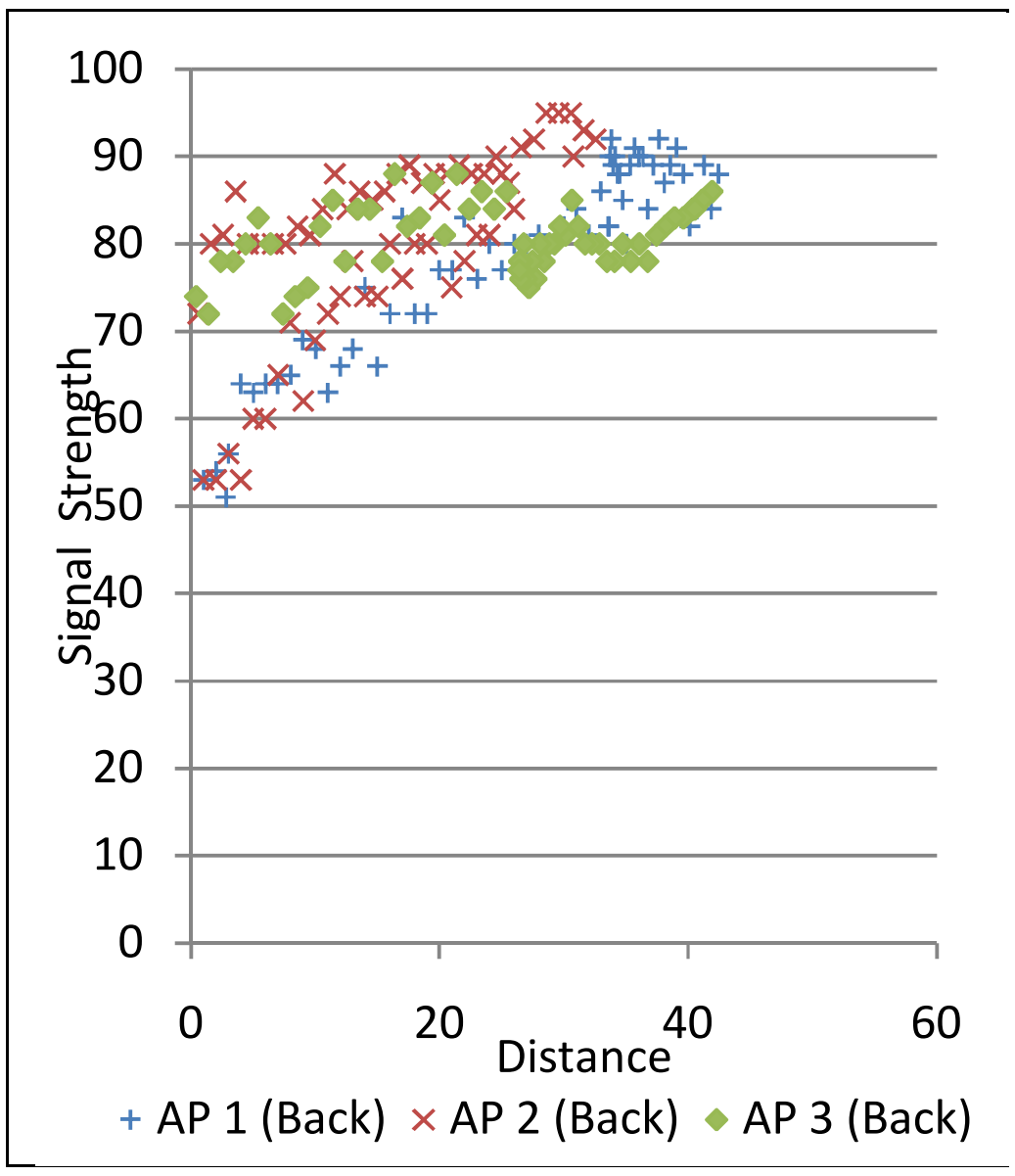

**Figure 13.** Distance and RSSI during Data Collection in Location 2 (User Orientation: $90^o$) (Note: Signal Strength in –dBm and Distance in meter).

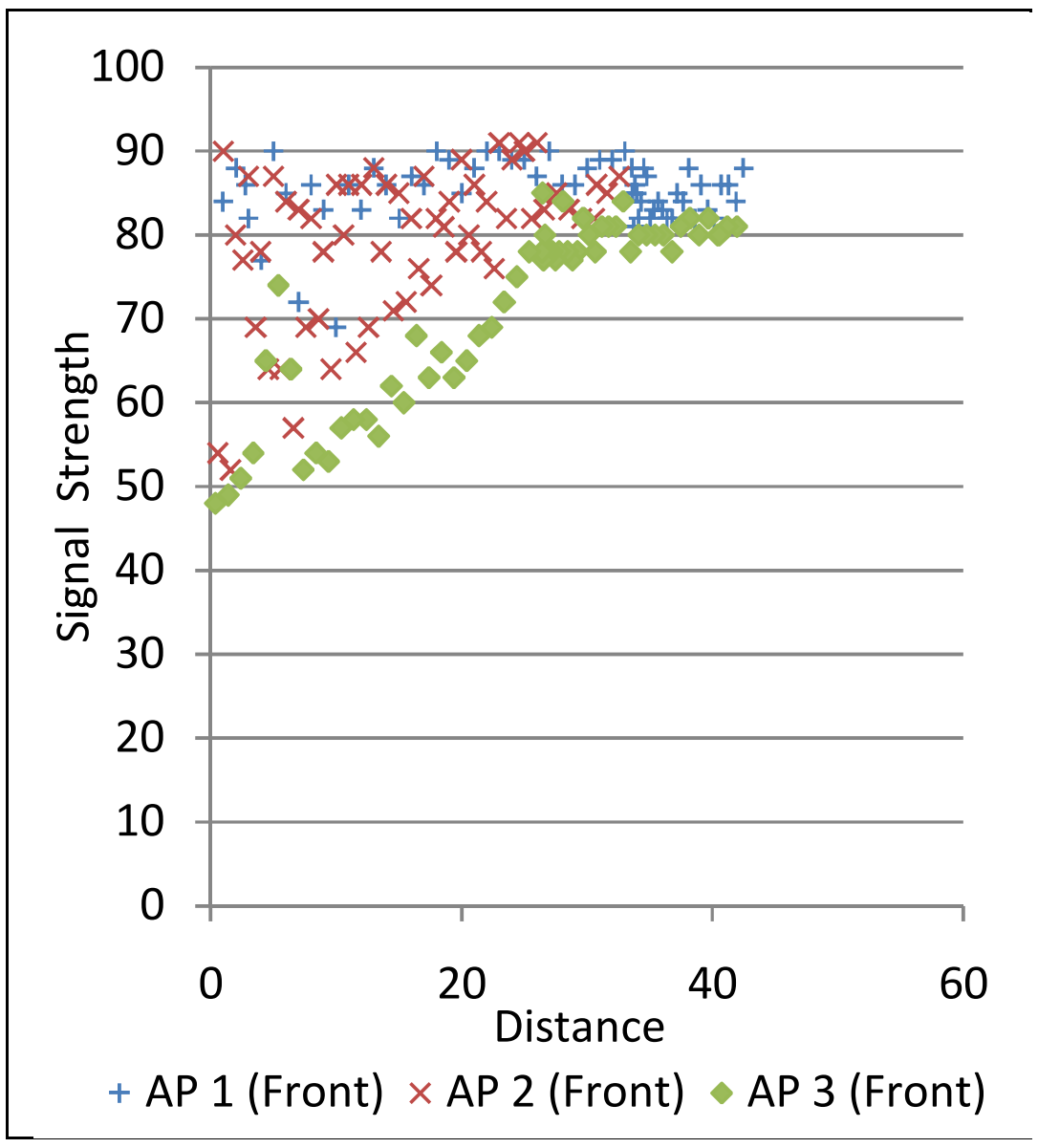

**Figure 14.** Distance and RSSI during Data Collection in Location 2 (User Orientation: $270^o$). (Note: Signal Strength in –dBm and Distance in meter).

Specifically in Figure 7, 8, 9 and 10 shows the signal strength value obtained by mobile device is different dependence on user orientation in the Location 1. For example, the signal strength value of AP2 is the most higher during user orientation: 270$^{o}$ (For detail, see Figure10) compared with signal strength obtained in 0$^{o}$ (For detail, see Figure 7). Mean while in location 2, we can see the signal strength value obtained by mobile device in Figure 11, 12, 13 and 14. It is also shows that the signal strength also depends on the user orientation although the distance between mobile device receiver and AP is close. In this case, the signal strength value of AP3 is the most higher during user orientation: 270$^{o}$ (For detail, see Figure 14) compared with signal strength obtained in 0$^{o}$ (in Figure 11). The reason behind this, the orientation of user may contribute of blockage, thus the signal strength that obtained by mobile device will be reduced. This reason also applied with other value in the graph.

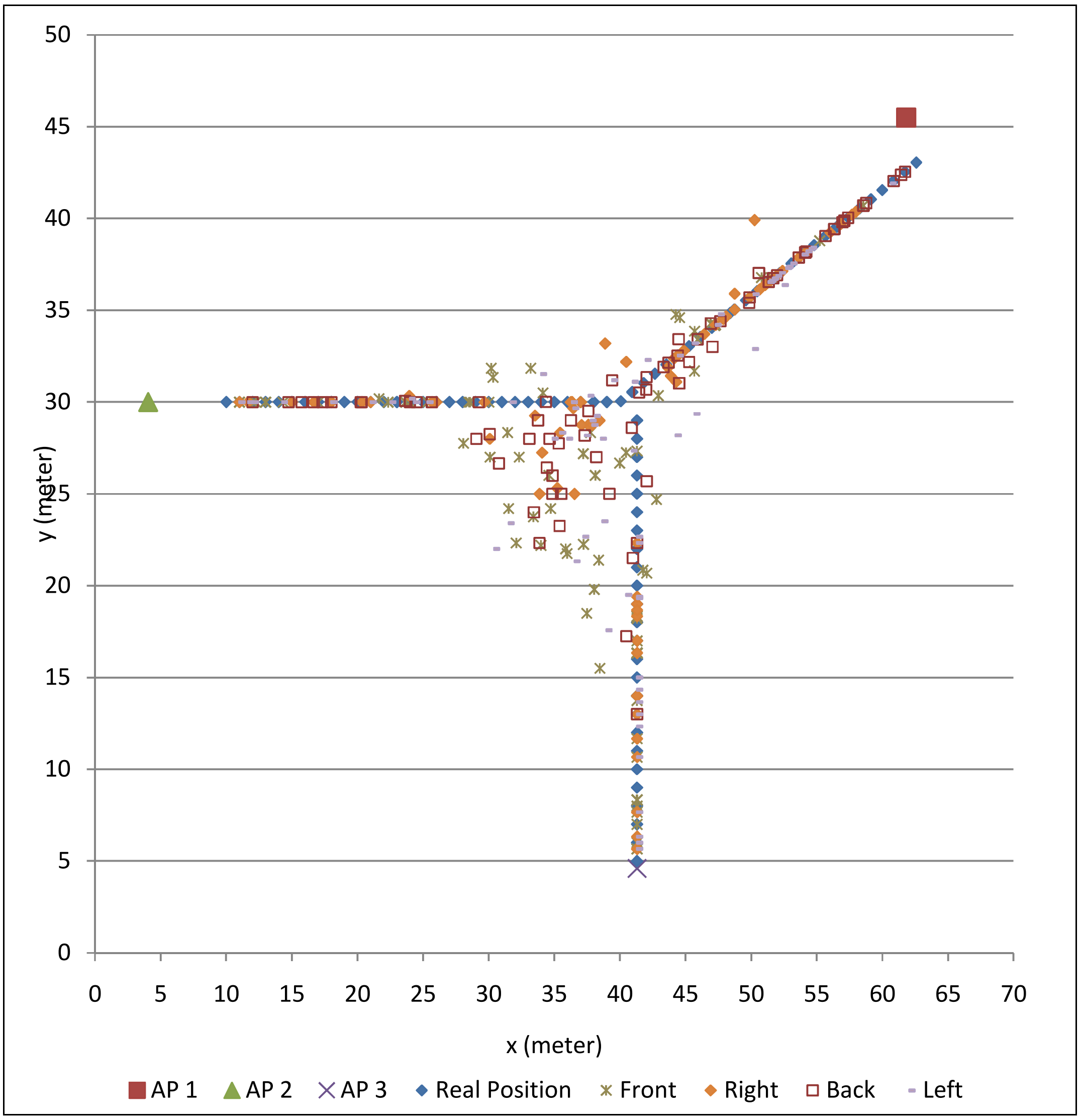

**Figure 15.** Localization Comparison Between Real Position And Experimented Position at Location 1. Front, Right, Back, and Left Refer to User Orientation during Experiment.

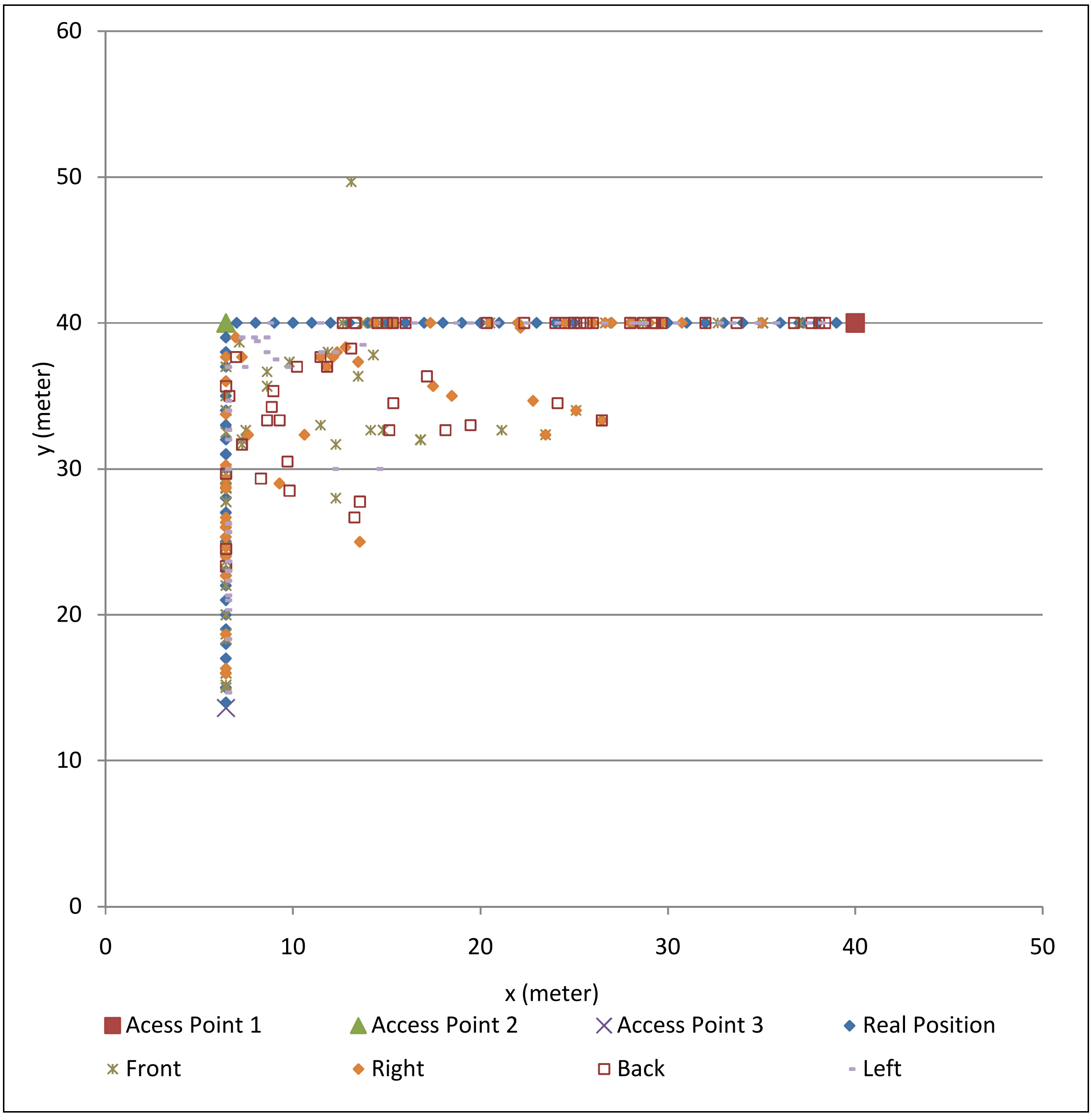

**Figure 16.** Localization Comparison Between Real Position And Experimented Position at Location 2. Front, Right, Back, and Left Refer to User Orientation during Experiment.

In overall, we also plot the overall result of WLAN positioning in Figure 15 (Location 1) and Figure 16 (Location 2). According to that figure, it is shows, most of the positioning information result getting error during WLAN mobile devices receiver located in the middle of location 1. The reason behind this is the signal strength that obtained in that area almost same, and it is make the searching algorithm (KNN nearest neighbor) easier to choosing unsuitable signal strength that refer to location in the database server (the signal strength that refer to the location was saved in the offline phase). However, the positioning information result getting better during WLAN mobile devices receiver is not located in the middle of location anymore. Its result the 37.35 percent of distance error at Location 1 was below 3.4 m. Besides that, 56.67 percent of distance error at Location 2 was below 2.9m. At the Location 1, As overall result, the average error of the indoor positioning in was 7.69777193 meter (Location 1) and 6.12233997 meter (Location 2).

## 5. Conclusion & Future Works

This paper discussed about our experiment for location determination by using hybridization between GPS and WLAN on mobile u-navigation system. The information from both of sensor output was switched based on type of environment (whether inside or outside building) in order to find the absolute of user target position. The result shows our proposed method can archive 7.69777193 meter (Location 1) and 6.12233997 meter (Location 2) during indoor positioning. Besides that, our proposed method also can survive during outdoor environment by archive 4.875m distance error. We believe our result is suitable and accepted for indoor/outdoor navigation system, although sometime the distance error of indoor environment slightly going too large.

As a future works, we will continue our experiment by using this result and combine with other mobile internal sensor such as camera in order to know how far our approach can be more ubiquity in many environments. Although the hybridization between camera (for example Augmented Reality technology) seems logic, but it is a challenging task for facility manager to manage the Augmented Reality (AR) tag in the building. Other issue will be raised such as facility management issue (this approach we will be elaborated in the next paper). The issue such as privacy is also need taken by developer; the implementation of the positioning algorithm in the standalone mobile client device is required. However, the implementation in standalone mobile device will be raising other issues such as performance since the development requires a huge location database. The intelligence of searching algorithm can be used to solve the performance issue. In that part, we believe it will be undertaken by the next researcher.

## 6. Acknowledgement

This paper was inspired from my master research project which is related to indoor positioning on mobile phone. The author also would like to thank our supervisor, Dr. Mohd Murtadha Mohamad and also our co-supervisor Dr. Maimunah Sapri for his/her insightful comments on earlier drafts of this paper. As additional, the author also would like to thank our research assistant officer for his insightful help in this experiment.